\documentclass[prb,twocolumn,showkeys]{revtex4}
\usepackage{graphicx}
\begin{document}
\title{Quantum dot from chiral metallic single walled nanotubes}
\author{R. Banerjea and J. Bhattacharjee \\
\textit{
School of Physical Sciences,\\
National Institute of Science Education and Research, \\
IOP Campus, Sachivalaya Marg, Bhubaneswar, India, 751005.}}
\begin{abstract}
We propose a simple approach to construct a quantum-dot and it's electrodes using chiral metallic 
single walled carbon nanotube (CM-SWCNT) segments of exactly opposite chiralities $(m,n)$ and $(n,m)$. 
The degree and energetics of electron confinement crucially depends on the choice of $n$ and $m$, and 
collinearity of the SWCNT segments constituting the quantum-dot and the electrodes.
All the segments can in principle be obtained through simple manipulations of fragments of a 
single nanotube of chirality either $(m,n)$ or $(n,m)$.
\end{abstract}
\maketitle
Graphene or carbon nanotube (CNT) based active elements have been long envisioned to have the potential to 
replace their silicon based counterparts and accelerate the pace of miniaturization of electronic devices. 
Such an optimism primarily stems from the fact that CNTs  can be either
semiconducting or metallic depending on their periodic structure, and can also endure large tensile strain.
But such a prospect has been difficult to realize due to absence of any controllable 
strategy to synthesize key components like, graphene sheets with regular edges or
CNTs of specific chiralities. 
Further, fabrication of active elements like diodes, transistors, quantum dots(QD) etc., 
requires synthesis of heterojunctions(HJ) made of CNTs of dissimilar chiralities. 
Not withstanding recent reports of chance occurrence of such HJs\cite{lieber1,lieberRev}, 
partial success in their direct synthesis\cite{expt1} and measurement of their nonlinear behavior
\cite{dekker}, which has inspired detailed theoretical 
studies\cite{ywoo,chico2,chico1,chico3,kawazoe,nardelli} of CNT-HJs as active elements, 
such exotic HJs are yet to be available at a scale appropriate for large-scale implementation of
CNT-HJ based devices. An alternate approach could be to construct devices made of segments 
which can be obtained through precisely controllable manipulations of fragments sourced from of a 
single but thoroughly characterized CNT.  With the advent of
Rayleigh scattering\cite{feng} based technique for determination of chirality and simultaneous measurement
of electrical conductivity\cite{qdotbhup}, it is now indeed possible to unambiguously characterize a SWCNT,
paving the way for device fabrication with such a CNT as the starting point.

In this work we propose construction of a QD-electrode assembly where the QD region and the two electrodes
on it's two sides, are constituted by three segments of chiral metallic SWCNTs with exactly opposite 
chiralities,  differing, in effect,  only in the sense of their helicities, which does not alter the 
electronic structure. 
We find that, if three such segments are connected longitudinally such that their axes are collinear and 
the chirality of the segment in the middle is opposite to those of the segments on it's two sides, 
then the segment in the middle can strongly confine electronic states with regularly spaced energies, 
if the chiralities are suitably chosen.

\begin{figure}[b]
\centering
\includegraphics[scale=0.30]{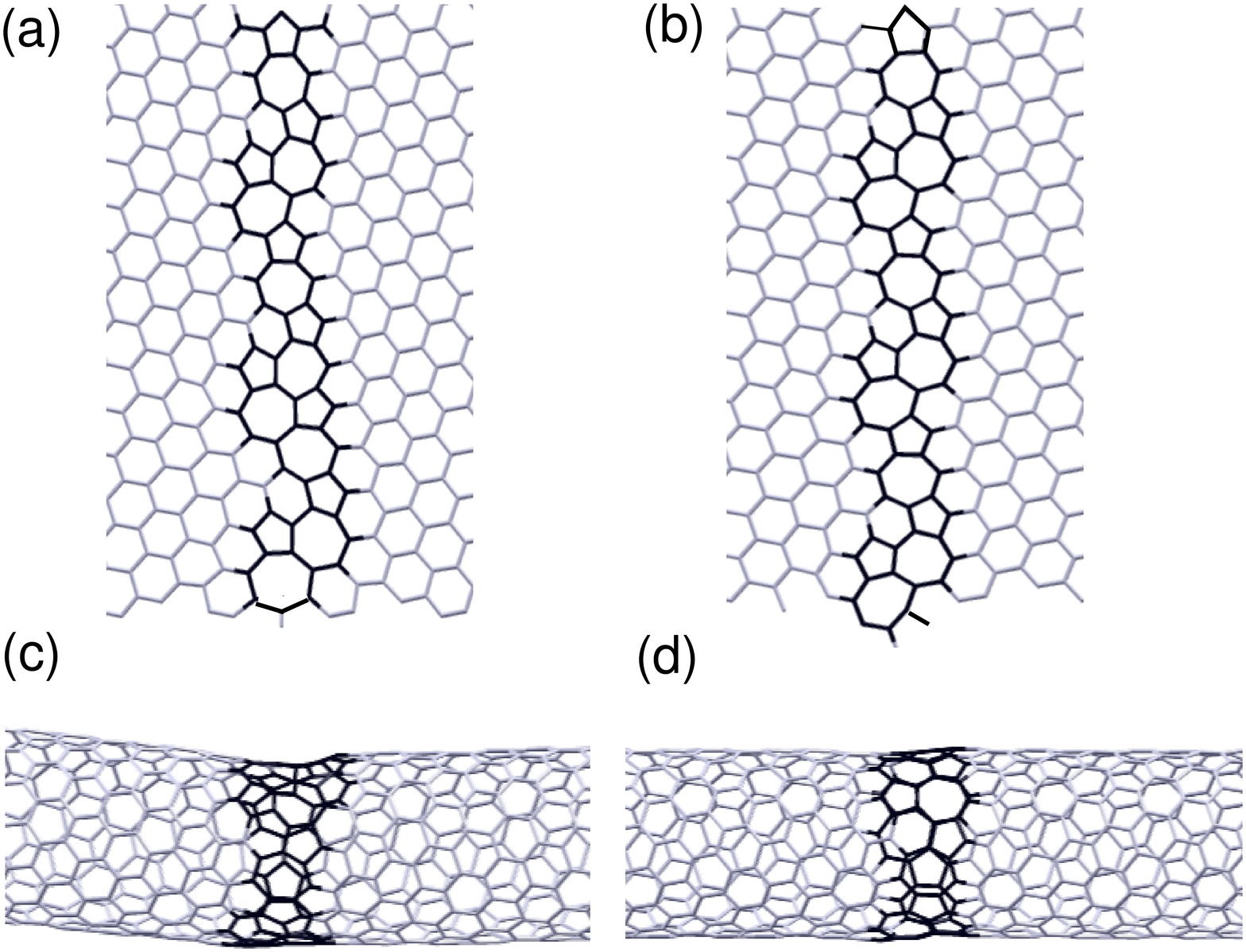}
\caption{(a) Non-uniform and (b) uniform distribution of defects (shown in black lines) at the (12,3)(3,12) interface. 
(c-d) Corresponding energy minimized structures of the resultant HJs obtained using Tersoff-Brenner \cite{tb1,tb2} 
classical force fields.}
\label{shape}
\end{figure}

\begin{figure*}[t]
\centering
\includegraphics[scale=0.55]{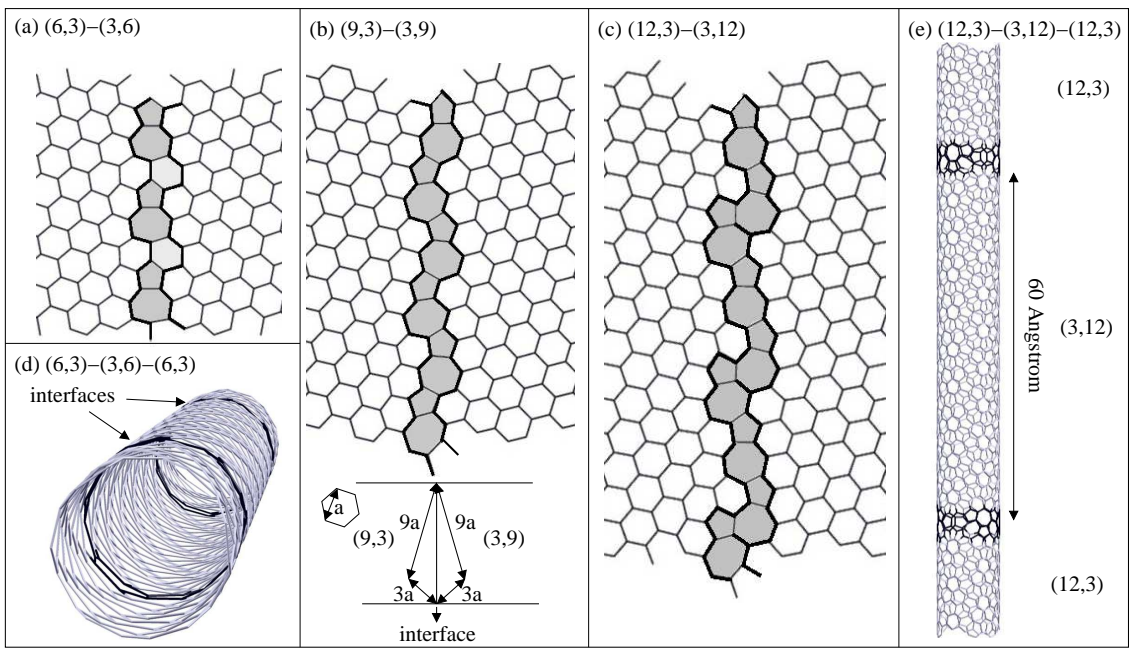}
\caption{(a-c): Distribution of defects (shown in darker shed) at the interfaces leading to linear HJ. (d-e)
resultant HJ assemblies with two such interfaces, one inverted with respect to the other.}
\label{interface}
\end{figure*}

We choose three sets of opposite chiralities [(6,3),(3,6)], [(9,3),(3,9) and [(12,3),(3,12)] to demonstrate
the possibility of the type of QDs proposed in this work. For the three chosen sets of chiralities, 
each of the two interfaces in the corresponding QD-electrode 
assemblies require three, six and nine pentagon-heptagon defect pairs\cite{jbjn} respectively, 
in order for the QD-electrode assemblies to be linear in shape.
It is expedient to mention that in general, any two SWCNTs can be connected\cite{dunlap,contub} 
by just one pentagon and one heptagon, but such heterojunctions are unavoidably bent at the interface.
Notable that pentagons and heptagons are the two simplest topological defects possible in graphene.
Evidently, with the increase of $|m-n|$ the number of pentagon-heptagon defect pairs required at the 
interfaces to form linear HJs increases. 
Within each pair we consider the pentagon and the heptagon adjacent to each other, 
sharing a common C-C bond. 
Such a paired arrangement, we refer here as (5-7), minimizes the net strain due to the individual defects. 
The defect pairs can in principle be arranged in a large variety of ways at the interfaces and each arrangement 
would result into different values of net strain, total energy and degree of linearity of shape of the 
resultant HJs. 
Maximally uniform distribution of defect pairs at the interface leads to maximally linear resultant HJ\cite{jbjn}  
due to uniform distribution of strain over the circumference at the interface between two SWCNTs. 
As evident in Fig-\ref{shape}, linearity of shape of such HJs strictly require uniformity of 
circumferential distribution of defects at the interface. 
This implies that if the QD-electrode assemblies are mechanically
constrained to have a linear shape, then the defects at the interfaces between the SWCNT 
segments will naturally occur arranged at the interfaces in a maximally uniform pattern.
Such a uniform distribution of defects at the two interfaces in a QD-electrode assembly can be
easily maintained if the open ends of the electrodes  are firmly attached to a suitable substrate or
a measurement probe.
This implies an additional and rather realistic constraint of fixed linear distance between the 
two ends of a QD-electrode assembly. This ensures that even at high enough temperatures, 
the uniform nature of defect distribution at the two interfaces in a QD-electrode assembly, 
will not degenerate because any deviation from the uniform distribution
will tend to bend the HJ at the interfaces, 
amounting in effect to stretching of the constituent SWCNT segments, and thus will be 
thermodynamically unfavourable.
Unwound interfaces of linear SWCNT HJs are shown in 
Fig-\ref{interface}-(a-c) for sets of opposite chiralities 
[(6,3),(3,6)], [(9,3),(3,9) and [(12,3),(3,12)].
Resultant HJ assemblies (6,3)-(3,6)-(6,3) and (12,3)-(3,12)-(12,3) are shown in Fig-\ref{interface}-(d) and (e) 
respectively.
\begin{figure*}[t]
\centering
(a)  \hspace{8cm} (b)\\
\includegraphics[scale=0.55]{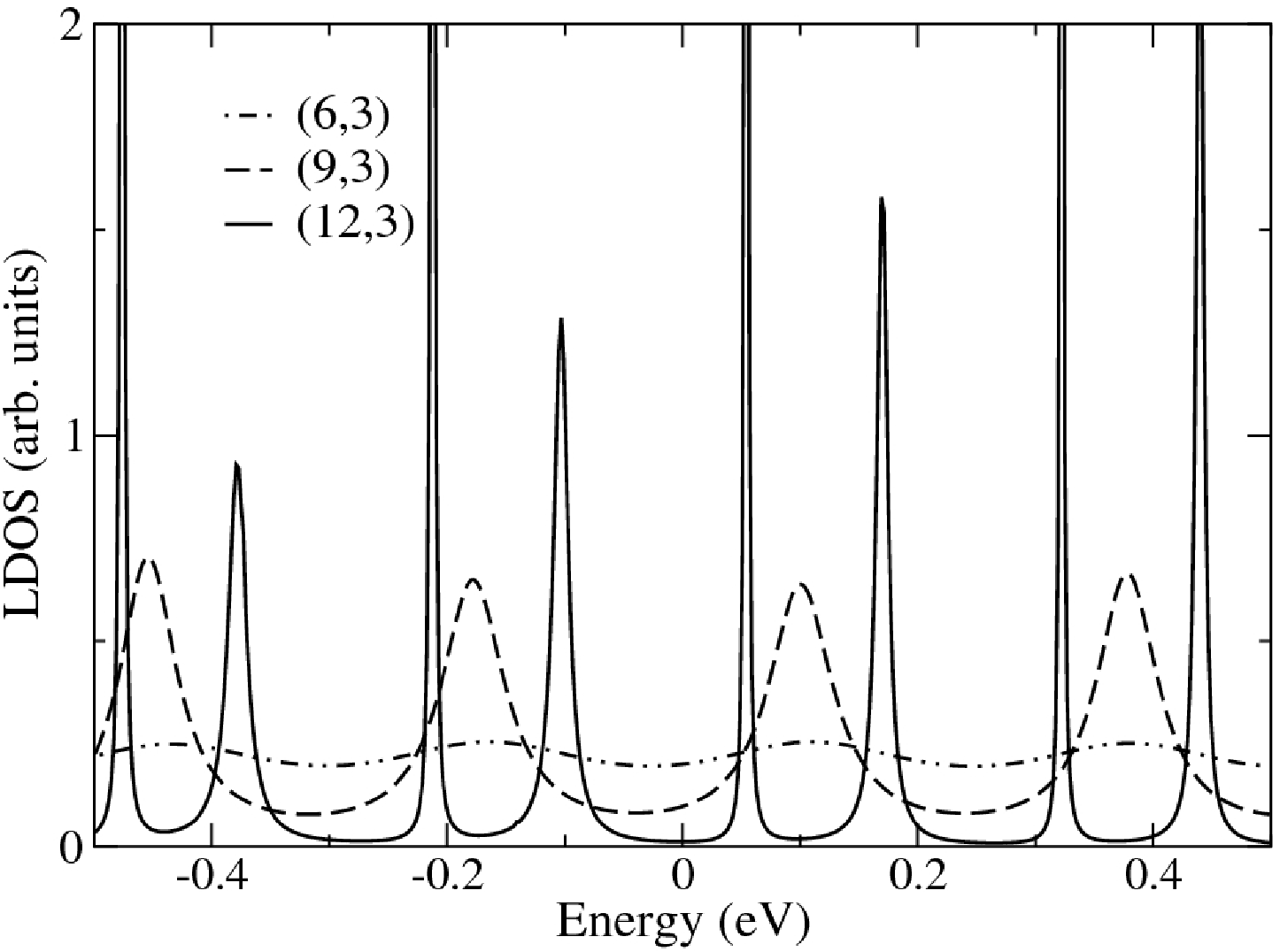}\hspace{1cm}
\includegraphics[scale=0.55]{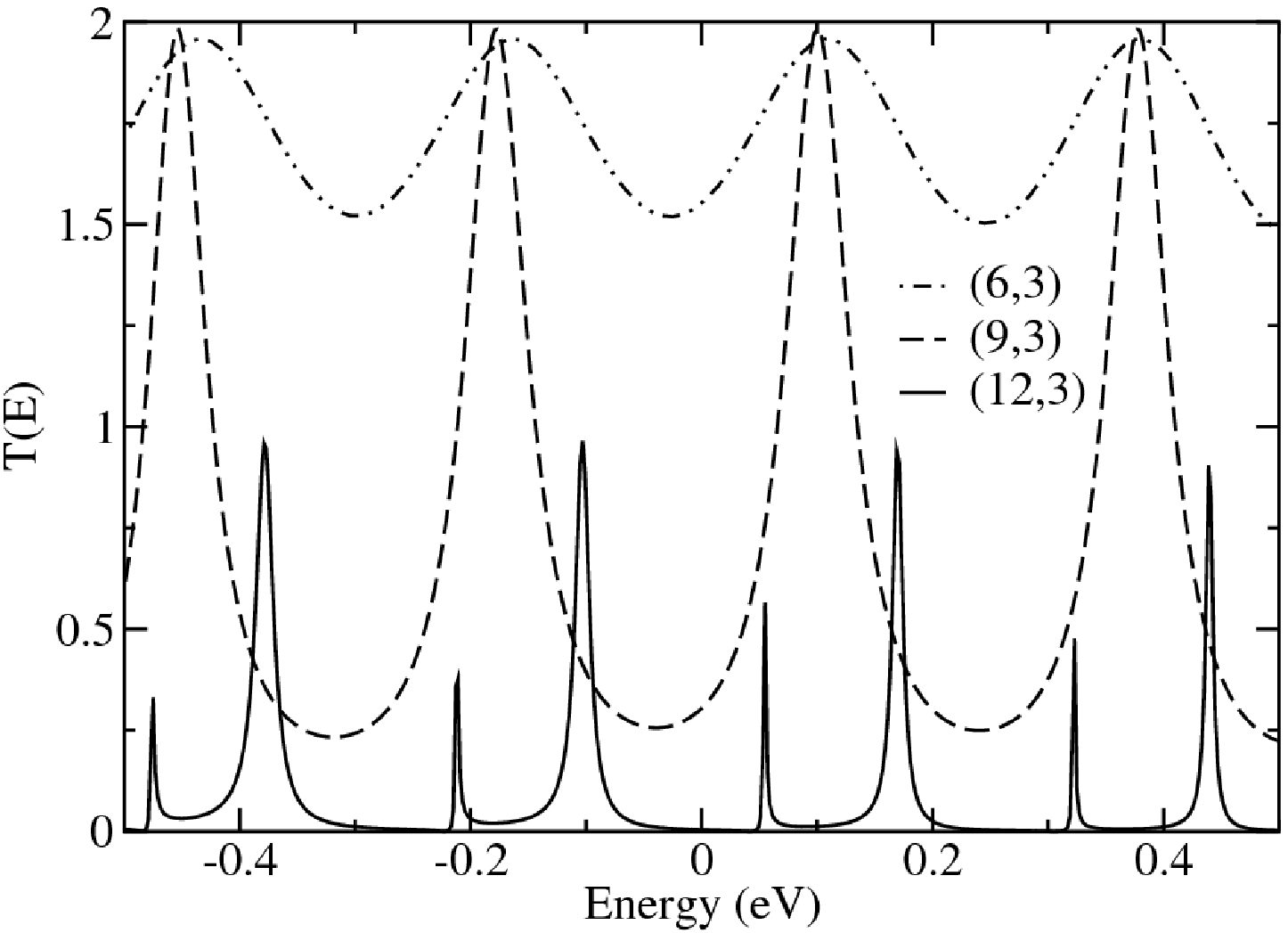}
\caption{(a) local density of states (LDOS) for a small region in the middle of the channel and (b) transmission function (T(E)) for the (n,m)-(m,n)-(n,m) HJs. }
\label{ldos}
\end{figure*}

The local density of states(LDOS) and conductance $T(E)$ at zero bias are calculated using the surface 
Green's function matching method\cite{gf1}, wherein, between two electrode regions at left (L) and right (R), 
a channel region (C) is considered to contain inside the possible active device. 
In the present case, the active devise consists of the 
electron confining CM-SWCNT segment and the two interfaces connecting it to the other two CM-SWCNT
segments. A large enough channel extending into the bulk of the CM-SWCNTs segments on two sides allows 
the Green's function associated with the channel to be correctly represented by\cite{dutta}
$G(E)=\left(E I - H_C - \Sigma_L - \Sigma_R \right)$, where $\Sigma_L=\tau_{CL} g_L \tau_{CL}^\dagger $ 
and $ \Sigma_R=\tau_{CR} g_R \tau_{CR}^\dagger$ are self energy contributions of the electrodes to the channel.
The surface Green's function $g$ of the electrodes are calculated recursively\cite{gf2} while $\tau$ is the coupling 
between the channel and the electrodes. 
 We consider nearest neighbor $\pi$ orbital tight-binding Hamiltonian with a hopping integral of value 
$-2.7eV$\cite{tparam}. The LDOS and transmission function are calculated as\cite{dutta}
$D\left(E\right)=\mbox{Trace}\left[\mbox{Imag}\left\{G\left(E\right)\right\}\right]/2\pi$ and 
$T\left(E\right)=\mbox{Trace}\left[\Gamma_L G \Gamma_R G^\dagger\right] $ respectively, where 
$\Gamma=\mbox{Imag}\left(\Sigma\right)$ are the broadening matrices for the channel due to coupling with the electrodes.
   
Local symmetry at the 5-7 defect sites being different than that in the bulk of graphene or CNT, increase
in number of 5-7 defects at interfaces reduces inter-segment mixing between electronic states in the 
segments that the interfaces connect. With increasing defect densities, the interfaces at the two ends
thus increasingly confine electronic states within the CM-SWCNT segment in the middle, constituting essentially a 
quantum dot(QD). Expectedly, the width of the LDOS peaks (Fig-\ref{ldos}-(a)) and the corresponding transmission peaks 
(Fig-\ref{ldos}-(b)) decreases as we go from QD-electrode assemblies made from SWCNTs 
[(6,3),(3,6)] to [(9,3),(3,9)] and then to [(12,3),(3,12)].
The energies of the confined states crucially depend on the nature of 
arrangement of defects at the interfaces. Advantageously, maximally uniform distribution of defects at the interfaces
leads to confined states 
regularly spaced in energy, as evident from energies associated with the LDOS peaks in Fig-\ref{ldos}-(a). 
Such states simply result from regular discretization of the lowest conduction band and 
the highest valence band, both of which have perfectly linear dispersion within the simplest tight-binding model 
considered in this work, which  does not take into account the effects of curvature.
Through systematically increasing $|m-n|$ we find that if $m/n > 3$, as it is for (12,3), the resultant QD-electrode 
assemblies, if linear in shape, consists of enough defects at the interfaces, resulting into substantial confinement of
electronic states within the QD segment. 
Thus in general, ensuring the CM-SWCNT based QD-electrode assemblies to be linear in shape
and $m/n > 3$, naturally increases the probability for substantial confinement and regular spacing of energy 
of the confined states.

Non uniform distribution of defects at the two interfaces on the two sides of the QD segment results into 
different lengths of longitudinal confinement at different circumferential  regions of the QD segment 
and hence different discretization of the lowest conduction band and the highest valence band. 
This leads to increase in number of confined states with irregularly spaced energies and different degrees of 
longitudinal localization over different circumferential regions of the QD segment.
Fortunately such a scenario is easily avoided by imposing the mechanical constraint
of linearity of shape of the  QD-electrode assembly and thereby ensuring uniform distribution of defects at the
interfaces.

Interestingly, it may be possible to obtain all the three components of a QD-electrode assembly, ie.,
the two electrodes and the QD segment in the middle, from a single long enough CM-SWCNT. 
After selecting out an accurately characterised CM-SWCNT of a suitable chirality as dicussed above, the tube 
may be fragmented into three, such that the fragment in the middle is of length equal 
to that of the desired length of electron confinement. The fragment in the middle, can be unzipped\cite{tubetorib} 
to form a ribbon.
Then through controlled hydrogen adsorption, the ribbon can possibly be again zipped back \cite{ribtotube}
to form a nanotube, in the direction opposite to that in which it was unzipped.
It can be verified easily that the chirality of the resultant tube will be exactly opposite to that of the 
original tube. Finally, through high temperature annealing\cite{ijima2}, the segments can be connected 
back through seamless interfaces to form the QD-electrode assembly.
To perform such an annealing which does not leave any
dangling bond at the interface, in each of the three SWCNT segments, the ends which are to be 
connected,  might require to be 
capped\cite{cap} prior to annealing, as demonstrated in \cite{ijima2}. 
During the annealing process the capped ends of the segments 
may be brought effectively in contact with each other and held such that their axes are collinear.
Finally, two ends of the resultant QD-electrode assembly can be attached on to suitable substrates or 
measuring probes, like AFM tips.

In conclusion, we have introduced a new approach to construct carbon nanotube based 
active devices whose components can be obtained from of a single well characterized tube.
We have demonstrated such an approach by suggesting construction of a robust quantum dot device by assembling
chiral metallic single walled carbon nanotube segments of opposite chiralities. The segments 
can be obtained from fragments of a single such nanotube through a sequence of
chemical and mechanical manipulations, each of which has been already reported individually in recent years.

In acknowledgement, RB thanks the Dept. of Science and Technology of the Govt. of India for his
INSPIRE fellowship. The authors thank NISER, Inst. of Phys. and the Dept. of Atomic Energy of the Govt. of India
for generous support.


\begin{references}
\bibitem{lieber1} M. Ouyang, J.-L. Huang, C. L. Cheung, C. M. Lieber, Science \textbf{291}, 97 (2001).
\bibitem{lieberRev} M. Ouyang, J.-L. Huang, C. M. Lieber, Acc. Chem. Res. \textbf{35}, 1018 (2002).
\bibitem{expt1} Y. Yao, Q. Li, J. Zhang, R. Liu, L. Jiao, Y. T. Zhu, Z. Liu,  Nature Mat. \textbf{6}, 293 (2007).
\bibitem{dekker}Z. Yao, H. W.C. Postma, L. Balents, and C. Dekker, Nature \textbf{402}, 273 (1999)
\bibitem{ywoo} Hajin Kim, J. Lee, S.-J. Kahng, Y.-W. Son, S. B. Lee, C.-K. Lee, J. Ihm, and Young Kuk, Phys. Rev. Lett. \textbf{90}, 216107 (2003).
\bibitem{chico2} L. Chico, Lorin X. Benedict, Steven G. Louie, and Marvin L. Cohen, Phys. Rev. B \textbf{54}, 2600 (1996).
\bibitem{chico1}L. Chico, Vincent H. Crespi, Lorin X. Benedict, Steven G. Louie, and Marvin L. Cohen, Phys. Rev. Lett. \textbf{76}, 971 (1996).
\bibitem{chico3} L. Chico, M. P. López Sancho, and M. C. Muñoz, Phys. Rev. Lett. \textbf{81}, 1278 (1998).
\bibitem{kawazoe}A. A. Farajian, K. Esfarjani, and Y. Kawazoe, Phys. Rev. Lett. \textbf{82}, 5084 (1999)
\bibitem{nardelli} M. B. Nardelli, Phys. Rev. B \textbf{60}, 7828 (1999).
\bibitem{feng} M. Y. Sfeir, T. Beetz, F. Wang, L. M. Huang, X. M. H. Huang, M. Y. Huang, J. Hone, S. O'Brien, J. A. Misewich, T. F. Heinz, L. J. Wu, Y. M. Zhu, L. E. Brus, Science  \textbf{312},(5773) 554, (2006).
\bibitem{qdotbhup} B. Chandra, J. Bhattacharjee, M. Purewal, Y.-W. Son, Y. Wu, M. Huang, T. F. Heinz, P. Kim, J. B. Neaton, J. Hone, \textit{unpublished}.
\bibitem{jbjn} J. Bhattacharjee, J. B. Neaton, Unpublished.
\bibitem{dunlap} B. I. Dunlap, Phys. Rev. B \textbf{49}, 5643 (1994).
\bibitem{contub} S. Melchor, J. A. Dobado, J. Chem. Inf. Comput. Sci. \textbf{44}, 1639 (2004).
\bibitem{tb1} J. Tersoff, Phys. Rev. B  \textbf{37}, 6991 (1988). 
\bibitem{tb2} D. W. Brenner, Phys. Rev. B  \textbf{42}, 9458 (1990).
\bibitem{gf1} F. Garcia-Moliner, V. R. Velasco, Phys. Rep. \textbf{200}, 83 (1991).
\bibitem{dutta} S. Dutta, \textit{ Electron transport in Mesoscopic Systems}, Cambridge University Press: Cambridge (1995). 
\bibitem{gf2} M. P. Lopez-Sancho, J. M. Lopez-Sancho, J. Rubio,  J. Phys. F \textbf{14}, 1205 (1984).
\bibitem{tparam} X. Blasé, L. X. Benedict, E. L. Shirley, S. G. Louie, Phys. Rev. Lett. \textbf{72}, 1878 (1994).
\bibitem{tubetorib}Dmitry V. Kosynkin, Amanda L. Higginbotham, Alexander Sinitskii, Jay R. Lomeda, Ayrat Dimiev, 
B. Katherine Price,  James M. Tour, Nature \textbf{458}, 872 (2009). 
\bibitem{ribtotube}Decai Yu and Feng Liu, Nano Lett., \textbf{7}(10), 3046 (2007).
\bibitem{ijima2}C. Jin, K. Suenaga, S. Iijima, Nature Nanotechnology \textbf{3}, 17 (2008).
\bibitem{cap}N. de Jonge, M. Doytcheva, M. Allioux, M. Kaiser, S. A. M. Mentink, K. B. K. Teo, R. G. Lacerda, and 
W. I. Milne, Adv. Mater. \textbf{17}, No. 4, 451 (2005).
\end{references}
\end{document}